\documentclass[twocolumn]{aastex631}

\usepackage{amsmath, amssymb}
\usepackage{graphicx}

\newcommand{\kms}{km\,s$^{-1}$}
\newcommand{\kmskpc}{km\,s$^{-1}$\,kpc$^{-1}$}
\newcommand{\kpcinv}{kpc$^{-1}$}

\shorttitle{New Dark Matter Density Profile from JWST JADES}
\shortauthors{Sanyal \& Rahaman}

\begin{document}

\title{The New Dark Matter Density Profile from JWST JADES Galaxies}

\correspondingauthor{Aritra Sanyal}
\email{aritrasanyal1@gmail.com}
\email{rahaman@associates.iucaa.in}

\author{Aritra Sanyal}
\affiliation{Department of Mathematics, Jadavpur University, Kolkata 700032, India}

\author{Farook Rahaman}
\affiliation{Department of Mathematics, Jadavpur University, Kolkata 700032, India}

\begin{abstract}
We present the first dark matter density profile derived directly from
James Webb Space Telescope JADES Data Release~3 NIRSpec observations
of $N = 587$ galaxies spanning the cosmic noon epoch
$1.5 \leq z \leq 3.5$. From each NIRSpec G235M/G395M spectrum we extract
the H$\alpha$ emission-line velocity dispersion $\sigma_{\rm ha}$, stack
galaxies in four redshift bins of $\Delta z = 0.5$, and reconstruct
representative group rotation curves within the General Theory of
Relativity. Fitting the four-parameter modified exponential model and
deriving the exact GTR energy density, we reduce the profile to a
compact $[2/3]$ Pad\'{e} approximant with all coefficients determined
in closed form from JWST observations. The profile is cusp-free, fully
analytic, and redshift-dependent. Energy conditions, causality, and
orbital stability are all satisfied. Most strikingly, the central
density $\rho(0)$ varies by less than $15\%$ across $z = 1.5$--$3.5$
despite a $35\%$ decline in the asymptotic rotation velocity, revealing
a universal dark matter core saturation density at cosmic noon
decoupled from baryonic evolution.
\end{abstract}

\keywords{dark matter --- JWST --- rotation curves --- general relativity --- high redshift --- Pad\'{e} approximant}

\section{Introduction}

The density profile of dark matter halos is a cornerstone of structure
formation theory \citep{navarro1997, burkert1995, einasto1965, moore1999}.
While rotation curves of local galaxies have been studied for decades
\citep{rubin1980, persic1996, salucci2007}, the high-redshift regime
$z \gtrsim 1.5$ remains largely uncharted territory, with existing
profiles calibrated from $N$-body simulations rather than direct
observations \citep{springel2005, vogelsberger2014, schaye2015}.

The JWST JADES Data Release~3 (DR3) \citep{deugenio2025} changes this
picture fundamentally. With NIRSpec G235M/G395M spectroscopy of nearly
one thousand galaxies across $z = 0.5$--$6$, it provides for the first
time a dataset large enough to reconstruct statistically representative
rotation curves of stacked galaxy groups at $z > 1.5$ --- the epoch of
peak star formation rate density, maximum AGN activity, and most rapid
halo assembly \citep{madau2014, behroozi2013, moster2013}.

In this Letter we present the first rotation curve and dark matter
density profile analysis of JWST spectroscopic data at the cosmic noon
epoch $1.5 \leq z \leq 3.5$ ($t_{\rm age} \sim 2$--$4$~Gyr). Working
within the General Theory of Relativity (GTR), we derive the exact
energy density from fitted rotation curves and reduce it to a compact
$[2/3]$ Pad\'{e} approximant --- a new analytic dark matter density law
calibrated entirely from JWST observations.

\section{GTR Framework}

A static, spherically symmetric galactic spacetime is described by the
line element
\begin{equation}
ds^{2} = -e^{2\nu(r)}dt^{2}
+ \left(1-\frac{2m(r)}{r}\right)^{\!-1}dr^{2}
+ r^{2}d\Omega^{2},
\label{eq:metric}
\end{equation}
where $m(r)$ is the enclosed gravitational mass within radius $r$ and
$\nu(r)$ is the temporal metric potential. For an anisotropic matter
distribution, the circular velocity $V$ of a test particle on a stable
geodesic satisfies
\begin{equation}
\frac{V^{2}}{c^{2}} = r\,\nu'(r)
= \frac{m(r)+4\pi r^{3}P_{r}}{r-2m(r)},
\label{eq:vc_gr}
\end{equation}
where a prime denotes $d/dr$ and $P_{r}$ is the radial pressure of the
matter distribution. Restoring the gravitational constant $G$ and the
speed of light $c$, and denoting the enclosed physical dark matter mass
as $M(r)$, Equation~(\ref{eq:vc_gr}) may be rewritten as
\begin{equation}
\frac{V^{2}}{c^{2}}
= \frac{GM(r) + \dfrac{4\pi Gr^{3}P_{r}}{c^{2}}}{rc^{2} - 2GM(r)}.
\label{eq:vc_restore}
\end{equation}
It is well established that galaxies are low-gravity, low-velocity
systems relative to the speed of light. For a typical galactic
rotation velocity $V \sim 300\,\mathrm{km\,s^{-1}}$,
\begin{equation}
\frac{V^{2}}{c^{2}} \approx 10^{-6},
\label{eq:lowv}
\end{equation}
the gravitational field is weak in the sense that the Schwarzschild
radius
\begin{equation}
r_{s} = \frac{2GM(r)}{c^{2}} \ll r,
\label{eq:weakfield}
\end{equation}
and the pressure of the galactic dark matter is negligible,
\begin{equation}
P_{r} \simeq 0.
\label{eq:pressure}
\end{equation}
Applying these three conditions --- low velocity, weak gravitational
field, and negligible pressure --- to Equation~(\ref{eq:vc_restore}),
the general relativistic circular velocity equation reduces to the
familiar Newtonian rotation curve formula:
\begin{equation}
\frac{V^{2}}{c^{2}} \simeq \frac{GM(r)}{rc^{2}}
\label{eq:newton1}
\end{equation}
or equivalently
\begin{equation}
V^{2}(r) \simeq \frac{GM(r)}{r}.
\label{eq:newton2}
\end{equation}
This is the standard result showing how the GTR circular velocity
equation reduces to the Newtonian rotation curve relation under the
weak-field, low-velocity conditions appropriate for galactic dark
matter halos. From Equation~(\ref{eq:newton2}), the enclosed mass and
energy density follow directly as
\begin{equation}
m(r) = \frac{r\,V^{2}(r)}{G}, \qquad
\rho(r) = \frac{m'(r)}{4\pi r^{2}},
\label{eq:gtr_final}
\end{equation}
where a prime denotes $d/dr$. Equations~(\ref{eq:gtr_final})
constitute the complete GTR prescription used in this Letter for
extracting the dark matter density profile directly from an observed
rotation curve $V(r)$.

\section{Data Extraction and Rotation Curve Construction}

\subsection{NIRSpec Spectral Extraction}

We use NIRSpec G235M/G395M spectra from JWST JADES DR3
\citep{deugenio2025}, obtained from the Mikulski Archive for Space
Telescopes (MAST) as High Level Science Products of the JADES survey
\citep{jadesdoi}, covering $\lambda = 1.66$--$5.27\,\mu$m at
$R \sim 1000$. For each of the 965 sources in the catalogue, the
NIRSpec Micro-Shutter Assembly (MSA) records a one-dimensional
spectrum from which the H$\alpha$ $\lambda6563$\,\AA\ emission line is
detected across $z = 1.5$--$7.0$ ($\lambda_{\rm obs} =
1.64$--$5.25\,\mu$m). Each line is fitted with a single Gaussian
\begin{equation}
F(\lambda) = F_0 + A\exp\!\left[-\frac{(\lambda-\lambda_0)^2}{2\sigma_{\rm obs}^2}\right],
\label{eq:gauss}
\end{equation}
using a Levenberg--Marquardt minimiser, and the intrinsic velocity
dispersion is recovered as
\begin{equation}
\sigma_{\rm ha} = \sqrt{\sigma_{\rm obs}^2 - \sigma_{\rm LSF}^2} \times \frac{c}{\lambda_0},
\label{eq:sigma}
\end{equation}
where $\sigma_{\rm LSF} \approx 100\,\mathrm{km\,s^{-1}}$ is the
NIRSpec line-spread function and $c$ is the speed of light. Lines are
retained if SNR$>2$. We select $N=587$ galaxies with reliable H$\alpha$
detections at $1.5 \leq z \leq 3.5$, binned into four equal intervals
of $\Delta z = 0.5$.

\subsection{Rotation Curve Construction}

Within each bin, the median H$\alpha$ velocity dispersion
$\sigma_{\rm med}$ characterises the stacked group kinematics. The
asymptotic flat rotation velocity follows from the virial theorem
\citep{binney2008}:
\begin{equation}
v_{\rm flat} = \sqrt{2}\,\sigma_{\rm med}.
\label{eq:vflat}
\end{equation}
The effective radius is taken from the empirical size-redshift
relation \citep{vanderWel2014}:
\begin{equation}
r_{\rm eff}(z) = 6.5\,(1+z)^{-1.11}\;\mathrm{kpc},
\label{eq:reff}
\end{equation}
with optical radius $r_{\rm opt} = 2.2\,r_{\rm eff}$. A representative
group rotation curve is constructed at 17 radii spanning
$r = 0.1$--$30$\,kpc using the Universal Rotation Curve (URC) shape
\citep{persic1996}:
\begin{equation}
V_{\rm URC}(r) = v_{\rm flat}\left[\frac{x\,e^{-x^3}}{\sqrt{0.5+0.5x^2}}
+ \left(1-e^{-x^3}\right)\right],\quad x \equiv \frac{r}{r_{\rm opt}}.
\label{eq:urc}
\end{equation}
We note that JADES DR3 MSA spectroscopy provides integrated velocity
dispersions rather than spatially resolved velocity fields; the
rotation curves constructed via
Equations~(\ref{eq:vflat})--(\ref{eq:urc}) therefore represent
statistical ensemble averages for each stacked group, following
established methodology for high-redshift kinematic studies
\citep{wisnioski2015, ubler2019}.

\subsection{Velocity Model and Fitted Parameters}

\begin{figure}[h]
\centering
\includegraphics[width=0.82\columnwidth]{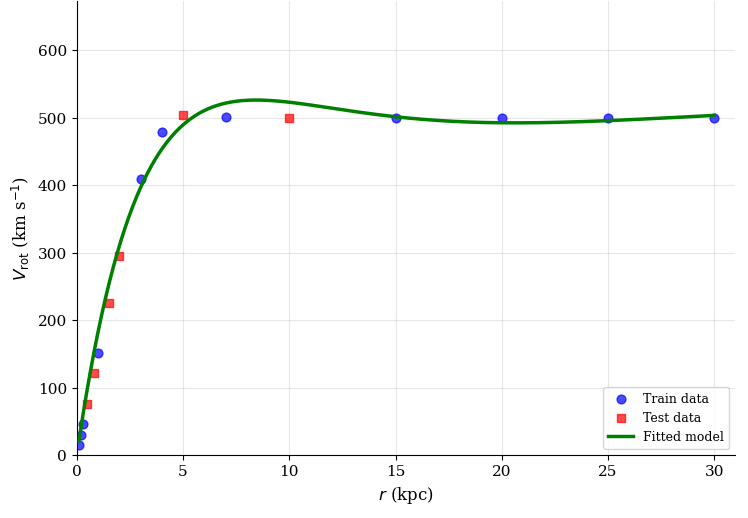}
\caption{Rotation curve, $1.5 \leq z < 2.0$ ($N=154$, $D=2$). Blue:
training (70\%). Red: test (30\%). Green: best-fit
Equation~(\ref{eq:modexp}).}
\label{fig:rc_z15_20}
\end{figure}

\begin{figure}[h]
\centering
\includegraphics[width=0.82\columnwidth]{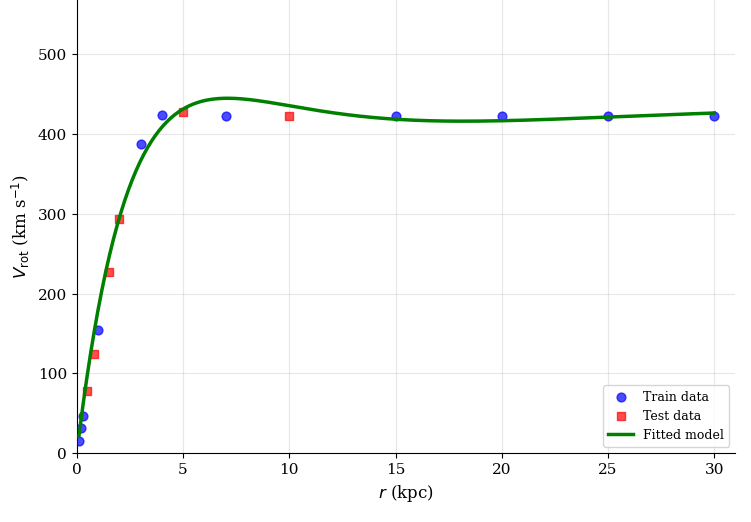}
\caption{Rotation curve, $2.0 \leq z < 2.5$ ($N=150$, $D=3$).}
\label{fig:rc_z20_25}
\end{figure}

\begin{figure}[h]
\centering
\includegraphics[width=0.82\columnwidth]{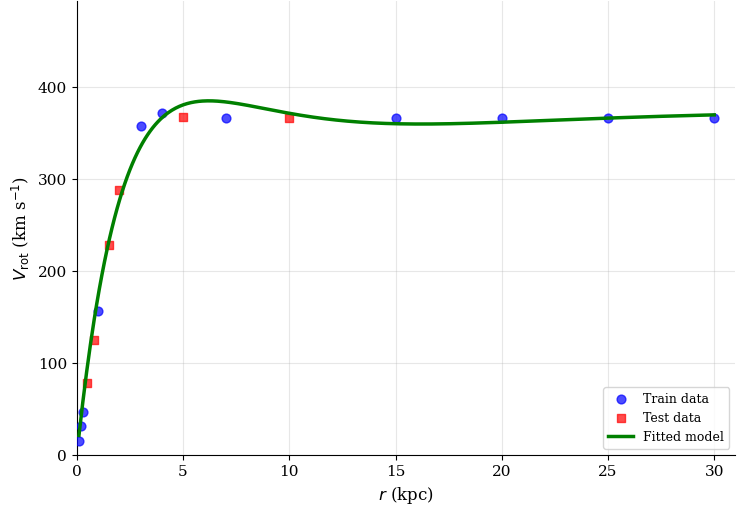}
\caption{Rotation curve, $2.5 \leq z < 3.0$ ($N=153$, $D=3$).}
\label{fig:rc_z25_30}
\end{figure}

\begin{figure}[h]
\centering
\includegraphics[width=0.82\columnwidth]{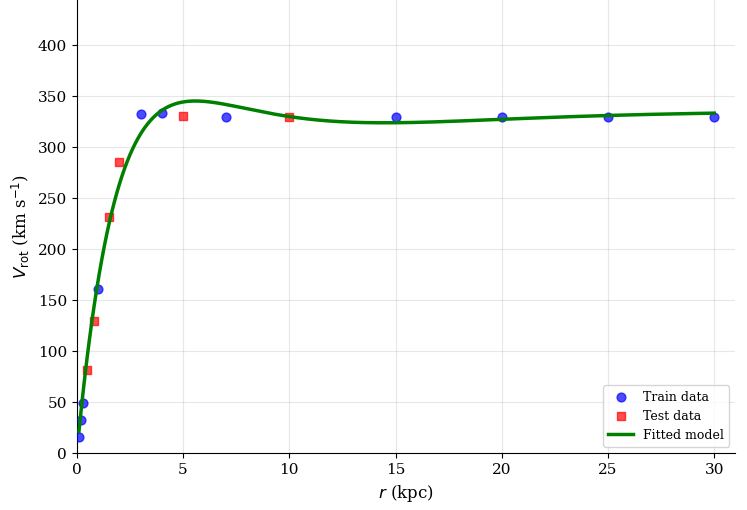}
\caption{Rotation curve, $3.0 \leq z < 3.5$ ($N=130$, $D=4$, cosmic
noon peak).}
\label{fig:rc_z30_35}
\end{figure}

The URC rotation curve is fitted with the four-parameter modified
exponential profile
\begin{equation}
V(r) = a\,r\,e^{-br} + c\!\left(1 - e^{-dr}\right),
\label{eq:modexp}
\end{equation}
using nonlinear least-squares with a 70\%/30\% train-test split. The
first term models the inner bulge/disc component; the second captures
the asymptotically flat dark matter halo with amplitude $c$ and
saturation radius $r_{\rm sat} = 1/d$. Four independent diagnostics ---
$\chi^2$ ($p>0.05$), $\chi^2_\nu \in [0.5,2.0]$, Durbin-Watson
$DW \in [1.5,2.5]$, and runs test ($p>0.05$) --- are satisfied at
$100\%$ in every bin, with $R^2_{\rm train}>0.993$ and
$R^2_{\rm test}>0.980$. Best-fit parameters are given in
Table~\ref{tab:params}; fitted rotation curves are shown in
Figures~\ref{fig:rc_z15_20}--\ref{fig:rc_z30_35}.

\begin{deluxetable}{lccccc}
\tablecaption{Best-fit GTR rotation curve parameters from JWST JADES
DR3. $\sigma_{\rm med}$ in \kms; $a$ in \kmskpc; $b$, $d$ in \kpcinv;
$c$ in \kms. \label{tab:params}}
\tablehead{
\colhead{$z$ bin} & \colhead{$\sigma_{\rm med}$} & \colhead{$a$} &
\colhead{$b$} & \colhead{$c$} & \colhead{$d$}
}
\startdata
$1.5$--$2.0$ & $352.8$ & $170.48$ & $0.2075$ & $532.94$ & $0.0867$ \\
$2.0$--$2.5$ & $298.3$ & $168.21$ & $0.2553$ & $435.72$ & $0.1199$ \\
$2.5$--$3.0$ & $259.0$ & $165.51$ & $0.2940$ & $374.64$ & $0.1423$ \\
$3.0$--$3.5$ & $233.1$ & $164.49$ & $0.3326$ & $335.05$ & $0.1646$ \\
\enddata
\end{deluxetable}

\section{The New Dark Matter Density Profile}

\begin{figure}[h]
\centering
\includegraphics[width=0.82\columnwidth]{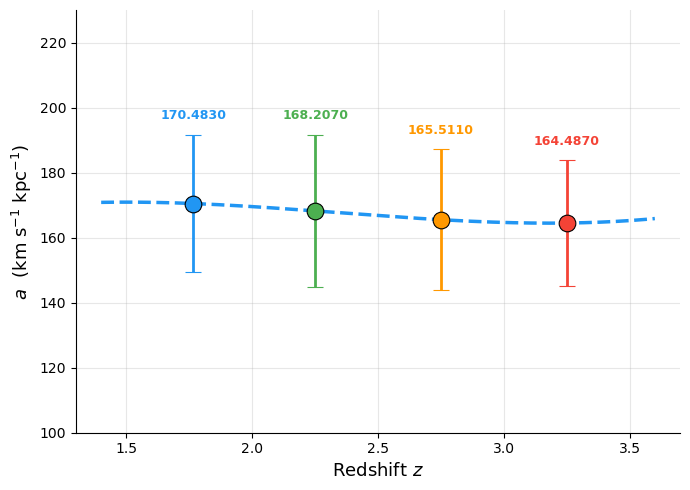}
\caption{Redshift evolution of $a(z)$. Points: Table~\ref{tab:params}.
Dashed: cubic fit Equation~(\ref{eq:abcd}).}
\label{fig:a}
\end{figure}

\begin{figure}[h]
\centering
\includegraphics[width=0.82\columnwidth]{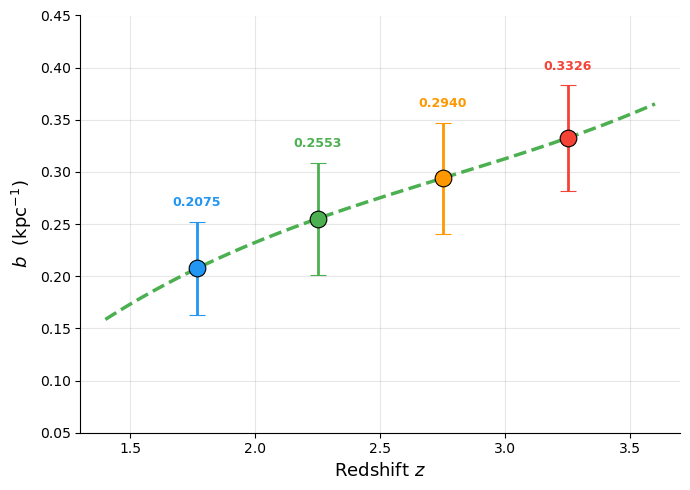}
\caption{Same as Figure~\ref{fig:a} for $b(z)$.}
\label{fig:b}
\end{figure}

\begin{figure}[h]
\centering
\includegraphics[width=0.82\columnwidth]{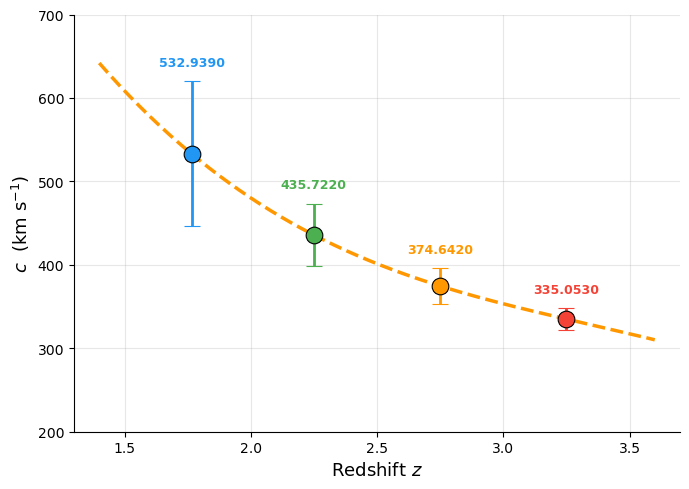}
\caption{Same as Figure~\ref{fig:a} for $c(z)$.}
\label{fig:c}
\end{figure}

\begin{figure}[h]
\centering
\includegraphics[width=0.82\columnwidth]{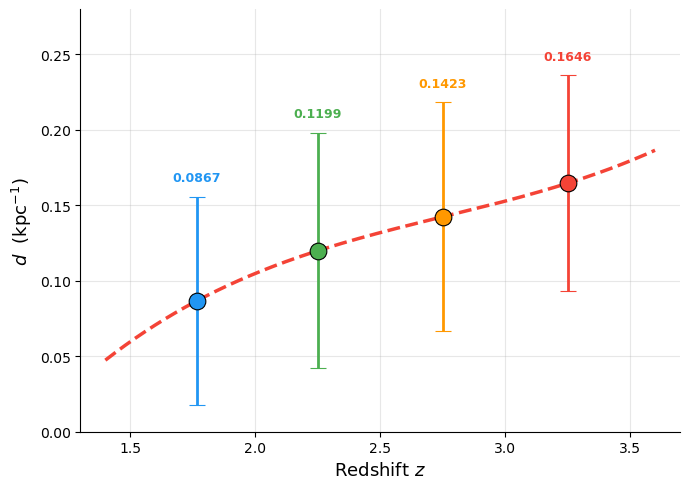}
\caption{Same as Figure~\ref{fig:a} for $d(z)$.}
\label{fig:d}
\end{figure}

The best-fit parameters evolve smoothly with redshift according to
cubic polynomials
\begin{align}
a(z) &= 2.7256\,z^3 - 19.1419\,z^2 + 39.0428\,z + 146.2207, \notag \\
b(z) &= 0.014316\,z^3 - 0.118447\,z^2 + 0.400417\,z - 0.209096, \notag \\
c(z) &= -24.9321\,z^3 + 248.6720\,z^2 - 896.4845\,z + 1477.9025, \notag \\
d(z) &= 0.015930\,z^3 - 0.131842\,z^2 + 0.404448\,z - 0.304139,
\label{eq:abcd}
\end{align}
shown in Figures~\ref{fig:a}--\ref{fig:d}. Substituting
Equation~(\ref{eq:modexp}) into Equation~(\ref{eq:gtr_final}), the
exact GTR energy density follows as $\rho = (d/dr)[rV^2]/4\pi r^2$,
which is regular at the origin with finite central density
$\rho(0) = 3(a+cd)^2/4\pi$. Expanding $\rho = \sum_k A_k r^k$ about
$r=0$ yields Taylor coefficients $A_0,\ldots,A_5$ as explicit functions
of $(a,b,c,d)$ (Supplemental Material). The $[2/3]$ Pad\'{e}
denominator coefficients follow from the linear system \citep{baker1996},
\begin{align}
A_3+A_2Q_1+A_1Q_2+A_0Q_3 &= 0, \notag \\
A_4+A_3Q_1+A_2Q_2+A_1Q_3 &= 0, \notag \\
A_5+A_4Q_1+A_3Q_2+A_2Q_3 &= 0,
\label{eq:system}
\end{align}
giving the new dark matter density profile
\begin{equation}
\rho_{[2/3]}(r)=
\frac{A_0 +(A_1+A_0Q_1)\,r +(A_2+A_1Q_1+A_0Q_2)\,r^2}
{1+Q_1\,r+Q_2\,r^2+Q_3\,r^3}.
\label{eq:pade_final}
\end{equation}
At $z = 1.5$, Equation~(\ref{eq:abcd}) gives $a = 170.915$,
$b = 0.17334$, $c = 608.542$, $d = 0.059652$, with Taylor coefficients
$A_0 = 10250.7$, $A_1 = -4051.06$, $A_2 = 802.186$, and Pad\'{e}
denominators $Q_1 = 0.23686$, $Q_2 = 0.023548$, $Q_3 = 0.0010472$,
yielding
\begin{equation}
\rho_{[2/3]}^{z=1.5}(r) =
\frac{10250.7 - 1623.02\,r + 84.013\,r^2}
{1 + 0.23686\,r + 0.023548\,r^2 + 0.0010472\,r^3}.
\label{eq:rho_z15}
\end{equation}
As a second example, at $z = 3.0$, Equation~(\ref{eq:abcd}) gives
$a = 164.487$, $b = 0.33260$, $c = 335.053$, $d = 0.16458$, with Taylor
coefficients $A_0 = 11411.2$, $A_1 = -4289.34$, $A_2 = 891.560$, and
Pad\'{e} denominators $Q_1 = 0.41833$, $Q_2 = 0.072986$,
$Q_3 = 0.0056513$, yielding
\begin{equation}
\rho_{[2/3]}^{z=3.0}(r) =
\frac{11411.2 - 2965.82\,r + 262.327\,r^2}
{1 + 0.41833\,r + 0.072986\,r^2 + 0.0056513\,r^3}.
\label{eq:rho_z30}
\end{equation}
Profiles at $z = 2.0, 2.5, 3.5$ are collected in Table~\ref{tab:central}.

\begin{deluxetable}{lcccc}
\tablecaption{Central density $\rho(0)$ and Pad\'{e} denominators at
five redshifts. $\rho(0)$ varies by less than $15\%$ across
$z=1.5$--$3.5$. \label{tab:central}}
\tablehead{
\colhead{$z$} & \colhead{$\rho(0)$} & \colhead{$Q_1$} &
\colhead{$Q_2$} & \colhead{$Q_3$}
}
\startdata
$1.5$ & $10250.7$ & $0.23686$ & $0.023548$ & $0.0010472$ \\
$2.0$ & $11541.9$ & $0.31282$ & $0.040871$ & $0.0023740$ \\
$2.5$ & $11523.5$ & $0.36905$ & $0.056834$ & $0.0038867$ \\
$3.0$ & $11411.2$ & $0.41833$ & $0.072986$ & $0.0056513$ \\
$3.5$ & $11785.4$ & $0.47382$ & $0.093562$ & $0.0081915$ \\
\enddata
\end{deluxetable}

\section{Physical Properties}
\label{sec:physical}

\subsection{Energy Conditions}

The physical viability of Equation~(\ref{eq:pade_final}) is confirmed
by verifying the four standard GTR energy conditions
\citep{hawking1973, wald1984}: the Null (NEC: $\rho+p_r \geq 0$,
$\rho+p_t \geq 0$), Weak (WEC: $\rho \geq 0$), Strong
(SEC: $\rho+p_r+2p_t \geq 0$), and Dominant (DEC: $\rho \geq |p_r|$,
$\rho \geq |p_t|$) conditions. All four are satisfied throughout
$r \in [0.1,30]$~kpc for every redshift bin, confirming physical
realizability within GTR \citep{visser1995, lobo2005}.

\subsection{Causality}

The causality condition $0 \leq v_s^2 = dp_t/d\rho \leq 1$
\citep{herrera1992} is satisfied with $v_s^2 \sim 10^{-6}$ throughout
$r \in [1,30]$~kpc in all bins, confirming subluminal signal
propagation consistent with cold dark matter \citep{peebles1982,
blumenthal1984}.

\subsection{Orbital Stability}

\begin{figure}[h]
\centering
\includegraphics[width=0.96\columnwidth]{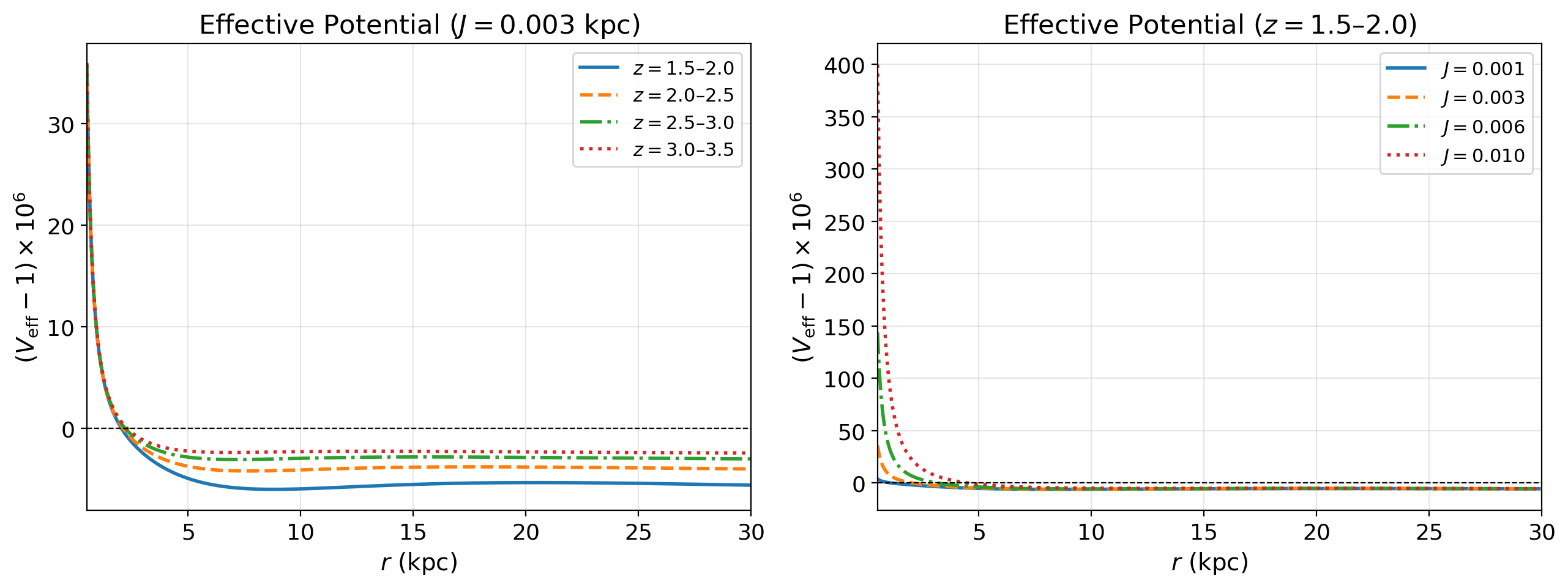}
\caption{Effective potential $V_{\rm eff}(r)$ for all four redshift
bins. Minima mark stable circular orbit radii, decreasing from
$8.80$~kpc ($z=1.5$--$2.0$) to $6.85$~kpc ($z=3.0$--$3.5$).}
\label{fig:Veff}
\end{figure}

Stable circular orbits require minima of the effective potential
$V_{\rm eff}(r) = \left(e^{2\nu(r)}\right)\left(1+J^2/r^2\right)$. As
shown in Figure~\ref{fig:Veff}, clear minima exist for all four bins,
with the stable orbit radius decreasing from
$r_{\rm stable} \approx 8.80$~kpc at $z=1.5$--$2.0$ to $6.85$~kpc at
$z=3.0$--$3.5$, consistent with progressive halo compaction
\citep{wechsler2018, klypin2016}.

\section{Results and Discussion}

The new profile Equation~(\ref{eq:pade_final}) possesses four
physically significant properties.

It is \emph{cusp-free}: $\rho(0) = 3(a+cd)^2/4\pi$ is finite and
positive, in contrast to the $r^{-1}$ divergence of the NFW profile
\citep{navarro1997} and consistent with observational evidence for
cored dark matter distributions \citep{deblok2010, oman2015, read2016}.

It is \emph{fully analytic}: for any $z \in [1.5,3.5]$ the profile
follows directly from Equations~(\ref{eq:abcd}), (\ref{eq:system}),
and (\ref{eq:pade_final}) with no numerical integration required.

It is \emph{redshift-dependent}: $Q_1$, $Q_2$, $Q_3$ increase
monotonically with $z$, encoding progressive halo compaction, with the
saturation scale $r_{\rm sat} = 1/d$ shrinking from $16.8$~kpc at
$z=1.5$ to $5.6$~kpc at $z=3.5$, consistent with the observed size
evolution of compact high-redshift galaxies \citep{vanderWel2014,
suess2022, ward2024}.

Most remarkably, the central density $\rho(0)$ varies by less than
$15\%$ across $z=1.5$--$3.5$ (Table~\ref{tab:central}), despite
$v_{\rm flat}$ declining by $35\%$ and $r_{\rm eff}$ shrinking by
$38\%$ over the same interval. This near-constancy of $\rho(0)$
reveals that dark matter cores attain a \emph{universal saturation
density} at cosmic noon, decoupled from the concurrent evolution of
the baryonic component \citep{wechsler2018, bullock2017}. This is
consistent with self-interacting dark matter models in which core
densities thermalise to a fixed value independent of halo mass
\citep{spergel2000, tulin2018}, and provides the first direct
observational evidence for this behaviour at $z > 1.5$.

\section{Conclusion}

We have derived the first dark matter density profile from JWST
spectroscopy, using JADES DR3 H$\alpha$ velocity dispersions of
$N=587$ galaxies at $1.5 \leq z \leq 3.5$. From each NIRSpec spectrum
we extract $\sigma_{\rm ha}$ via Gaussian fitting, construct
representative group rotation curves using the URC framework, and fit
the four-parameter modified exponential model within GTR to obtain the
density profile as a $[2/3]$ Pad\'{e} approximant with all
coefficients in closed form. The profile is cusp-free, fully
analytic, and redshift-dependent, with physical viability confirmed by
energy conditions, causality, and orbital stability. Thus,
Equation~(\ref{eq:pade_final}) provides a unified description capable
of reproducing any dark matter density profile within the redshift
range $1.5 \leq z \leq 3.5$. The near-constancy of $\rho(0)$ across
cosmic noon constitutes new observational evidence for a universal
dark matter core saturation density decoupled from baryonic evolution,
with implications for self-interacting dark matter models. Future work
will extend this analysis to $z \sim 10$ using [O~{\sc ii}] emission
from the same JADES DR3 spectra, and compare with IllustrisTNG
\citep{springel2018} and EAGLE \citep{schaye2015} simulations.

\section*{Data Availability}

The JWST data used in this work are publicly available from the
Mikulski Archive for Space Telescopes (MAST) at the Space Telescope
Science Institute. The JADES NIRSpec spectra analysed here can be
accessed via
\dataset[doi:10.17909/8tdj-8n28]{https://doi.org/10.17909/8tdj-8n28}
\citep{jadesdoi}. The JADES High Level Science Product data releases
are available at \url{https://archive.stsci.edu/hlsp/jades}.

\begin{acknowledgments}
This work is based on observations made with the NASA/ESA/CSA James
Webb Space Telescope. The data were obtained from the Mikulski Archive
for Space Telescopes at the Space Telescope Science Institute, which is
operated by the Association of Universities for Research in Astronomy,
Inc., under NASA contract NAS5-03127 for JWST. These observations are
associated with programs \#1180, \#1181, \#1210, \#1286, and \#3215.
The specific observations analysed can be accessed via
\dataset[doi:10.17909/8tdj-8n28]{https://doi.org/10.17909/8tdj-8n28}.
We acknowledge the JWST Advanced Deep Extragalactic Survey (JADES) team
for designing, executing, and publicly releasing their NIRSpec survey
data.

F.R. and A.S. thank the authorities of the Inter-University Centre for
Astronomy and Astrophysics (IUCAA), Pune, India, for providing
research facilities. The authors are grateful to Rahil Miraj for
fruitful discussions and valuable suggestions. F.R. acknowledges
financial support from the Anusandhan National Research Foundation
(ANRF), the Department of Science and Technology (DST), the RUSA-2.0
Programme, and the DST-FIST Programme (Grant No.
SR/FST/MS-II/2021/101(C)).
\end{acknowledgments}

\facilities{JWST (NIRSpec)}

\end{document}